\documentclass[sn-mathphys]{sn-jnl}

\jyear{2021}%

\theoremstyle{thmstyleone}%
\theoremstyle{thmstyletwo}%

\theoremstyle{thmstylethree}%

\begin{document}

\title[Slow slip detection with deep learning in multi-station geodetic time series]{Slow slip detection with deep learning in multi-station raw geodetic time series validated against tremors in Cascadia}

\author*[1]{\fnm{Giuseppe} \sur{Costantino}}\email{giuseppe.costantino@univ-grenoble-alpes.fr}

\author[1]{\fnm{Sophie} \sur{Giffard-Roisin}}

\author[1]{\fnm{Mathilde} \sur{Radiguet}}

\author[2,3]{\fnm{Mauro} \sur{Dalla Mura}}

\author[1]{\fnm{David} \sur{Marsan}}

\author[1]{\fnm{Anne} \sur{Socquet}}

\affil*[1]{\orgdiv{ISTerre}, \orgname{Univ. Grenoble Alpes, Univ. Savoie Mont Blanc, CNRS, IRD, Univ. Gustave Eiffel}, \orgaddress{\city{Grenoble}, \postcode{38000}, \country{France}}}

\affil[2]{\orgdiv{GIPSA-lab}, \orgname{Univ. Grenoble Alpes, CNRS, Grenoble INP}, \orgaddress{\city{Grenoble}, \postcode{38000}, \country{France}}}

\affil[3]{\orgname{Institut Universitaire de France (IUF)}, \orgaddress{\country{France}}}

\abstract{Slow slip events (SSEs) originate from a slow slippage on faults that lasts from a few days to years. A systematic and complete mapping of SSEs is key to characterizing the slip spectrum and understanding its link with coeval seismological signals. Yet, SSE catalogues are sparse and usually remain limited to the largest events, because the deformation transients are often concealed in the noise of the geodetic data. Here we present the first multi-station deep learning SSE detector applied blindly to multiple raw geodetic time series. Its power lies in an ultra-realistic synthetic training set, and in the combination of convolutional and attention-based neural networks. Applied to real data in Cascadia over the period 2007-2022, it detects 78 SSEs, that compare well to existing independent benchmarks: 87.5\% of previously catalogued SSEs are retrieved, each detection falling within a peak of tremor activity. Our method also provides useful proxies on the SSE duration and may help illuminate relationships between tremor chatter and the nucleation of the slow rupture. We find an average day-long time lag between the slow deformation and the tremor chatter both at a global- and local-temporal scale, suggesting that slow slip may drive the rupture of nearby small asperities.}




\maketitle
\section*{Introduction}

    Slow slip events (SSEs) generate episodic deformation that lasts from a few days to years. Like earthquakes, they originate from slip on faults but, unlike them, do not radiate energetic seismic waves. In the mid-1990s, Global Navigation Satellite System (GNSS) networks started to continuously monitor the ground displacement, providing evidence that SSEs are a major mechanism responsible for the release of stress in plate boundaries, as a complement to seismic rupture \cite{Dragert2001, Lowry2001, Schwartz2007, Ide2007, Mousavi2020}. This constituted a change of paradigm for the understanding of the earthquake cycle and of the mechanics of the fault interface. Twenty years later, the characterization of the full slip spectrum and the understanding of the link between slow slip and the associated seismological signals are hindered by our capacity to detect slow slip events in a systematic manner, more particularly those of low magnitude (typically lower than $M_w$ 6), even though a systematic and complete mapping of SSEs on faults is key for understanding the complex physical interactions between slow aseismic slip and earthquakes.
    Indeed, the small deformation transients associated with an SSE are often concealed in the noise \cite{Rousset2017, Frank2015}, making it difficult to precisely characterize the slip spectrum and provide fruitful insights into the fault mechanics \cite{Ide2007, Gomberg2016, Hawthorne2018}. Studies dealing with the detection and analysis of SSEs often rely on dedicated signal analysis, involving visual inspection of the data, data selection, denoising, filtering, geodetic expertise, dedicated modeling methods with a fine-tuning of the parameters, and also often complementary data such as tremor or LFE catalogs \cite{Frank2015, Frank2019, Michel2019, Bartlow2011, Radiguet2012}.

    The development of in-situ geophysical monitoring generates nowadays huge data sets, and machine learning techniques have been largely assimilated and used by the seismological community to improve earthquake detection and characterization \cite{Kong2019, Mousavi2020, Zhu2019, Woollam2022}, generating catalogs with unprecedented high quality \cite{Ross2019, Tan2021} and knowledge shifts \cite{Ross2020, Tan2020}. However, up to now, such techniques could not be successfully applied to the analysis of geodetic data and slow slip event detection because of two main reasons: (1) too few true labels exist to train machine learning-based methods, which we tackled by generating a realistic synthetic training data set, (2) the signal-to-noise ratio is extremely low in geodetic data \cite{rouet2021, costantino2022seismic}, meaning that we are at the limit of detection capacity. One possibility is to first pre-process the signals (via denoising, filtering, detrending), but this is at the cost of possibly corrupting the data. Instead, in this work, we assume that the information is already present in the raw time series and that our deep learning model should be able to learn the noise signature, and therefore to separate the noise from the relevant information (here, slow slip events). In order to develop an end-to-end model capable of dealing with raw geodetic measurements, it is necessary, on one hand, to set up advanced methods to generate realistic noise, taking into account the spatial correlation between stations as well as the large number of data gaps present in the GNSS time series. On the other hand, it involves developing a specific deep learning model able to treat multiple stations simultaneously, using a relevant spatial stacking of the signals (driven by our physics-based knowledge of the slow slip events) in addition to a temporal analysis. We address these two major drawbacks in our new approach and present SSEgenerator and SSEdetector, to our knowledge the first end-to-end deep learning-based detector, combining the spatiotemporal generation of synthetic GNSS time series containing modeled slow deformation (SSEgenerator), and a Convolutional Neural Network (CNN) and a Transformer neural network with an attention mechanism (SSEdetector), that proves effective in systematically detecting slow slip events in raw GNSS position time series from a large geodetic network containing more than 100 stations, both on synthetic and on real data.
    
\section*{Results}
    \subsection*{SSEgenerator: construction of the synthetic dataset}
        
        \begin{figure}
        \centering
            \includegraphics[width=1.\textwidth]{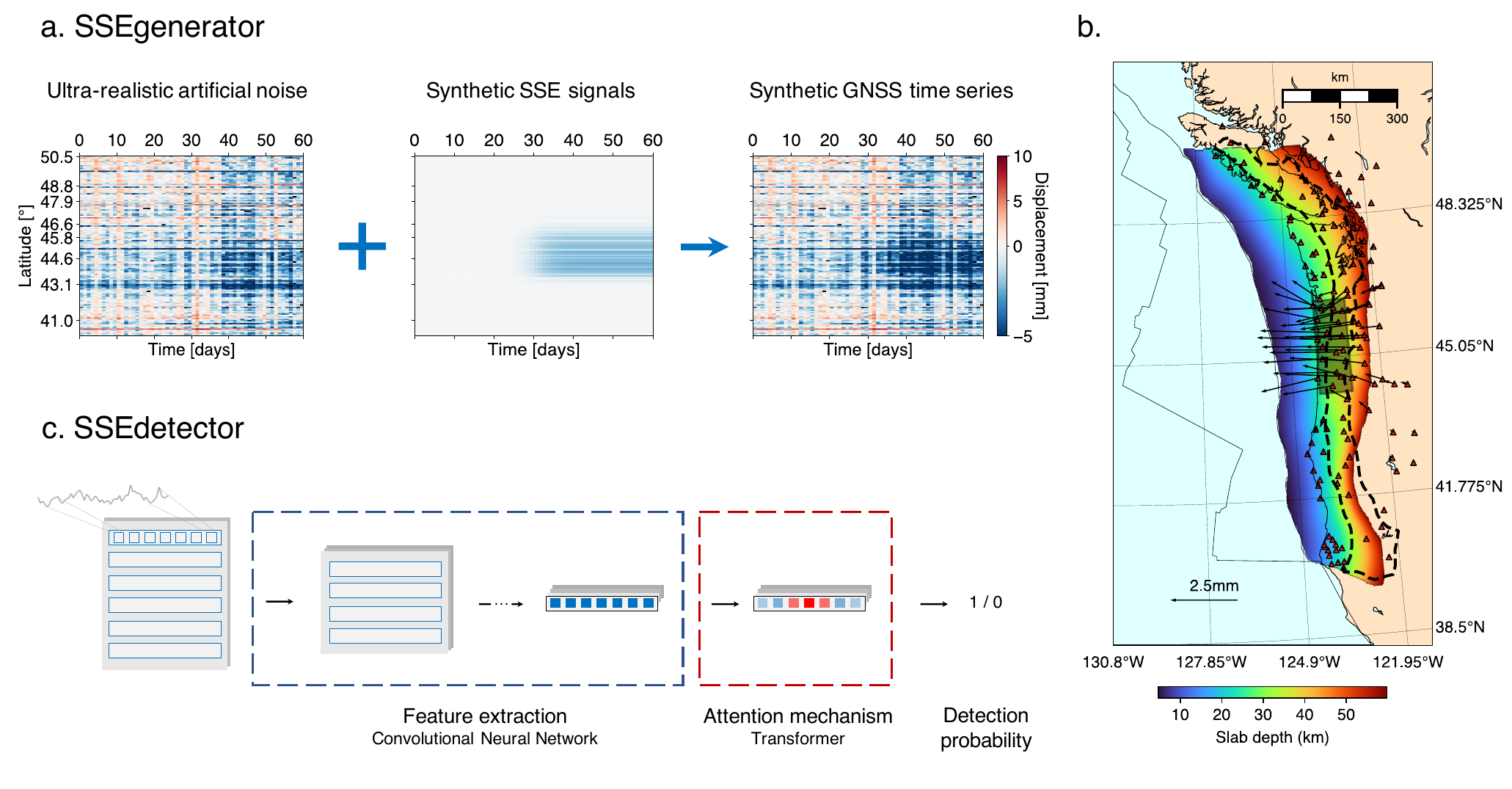}
            \caption{\textbf{Schematic architecture of SSEgenerator and SSEdetector.} (a) Overview of the synthetic data generation (SSEgenerator). In the matrix, each row represents the GNSS position time series for a given station, color-coded by the value of the position. The 135 GNSS stations considered in this study are here shown sorted by latitude. The synthetic static displacement model (cf. (b) panel), due to a $M_w \ 6.5$ event, at each station is convolved to a sigmoid to model the SSE transient, and is added to the ultra-realistic artificial noise to build synthetic GNSS time series. (b) Location of the GNSS stations of MAGNET network used in this study (red triangles). An example of synthetic dislocation is represented by the black rectangle, with arrows showing the modeled static displacement field. The heatmap indicates the locations of the synthetic ruptures considered in this study, color-coded by the slab depth. The dashed black contour represents the tremor locations from the PNSN catalog. (c) High-level representation of the architecture of SSEdetector. Input GNSS time series are first convolved in the time domain, then a higher weight is assigned to slow slip transients and a probability value is provided depending on whether slow deformation has been found in the data.}\label{fig1}
        \end{figure}

        We choose the Cascadia subduction zone as the target region because: (1) a link between slow deformation and tremor activity has been assessed \cite{Rogers2003} and a high-quality tremor catalog is available \cite{Wech2010}; (2) a preliminary catalog of SSEs has recently been proposed during the period 2007-2017 with conventional methods \cite{Michel2019}. This proposed catalog will be used for comparison and baseline for our results, which are expected to provide a more comprehensive catalog that will better show the link between slow deformation and tremors.

        To overcome the scarcity of catalogued SSEs, we train SSEdetector on synthetic data, consisting of simulated sets of geodetic time series for the full station network. Each set of signals (60 days and 135 stations) is considered as a single sample. In order to be able to detect SSEs in real raw time series, several characteristics need to be present in these synthetics. First, they must contain  a wide range of realistic background signals at the level of the GNSS network, \textit{i.e.} spatially and temporally-correlated realistic noise time series.  On the other hand, while half of the samples (\textit{negative} samples) will only consist of background noise, the other half must also include an SSE signal. For this, we modeled SSEs signals that are realistic enough compared to real transients of aseismic deformation. Finally, the synthetics should also carry realistic missing data recordings, as many GNSS stations have data gaps in practice.
        
        First, we thus generated ultra-realistic synthetic time series, that reproduce the spatial and temporal correlated noise of the data acquired by the GNSS network, based on the method developed by Costantino et al. \cite{costantino2022seismic}. This database of 60,000 synthetic time series was derived from real geodetic time series (details in Methods).
        We select data in the periods 2007-2014 and 2018-2022 as sources for the noise generation, while we keep data in the period 2014-2017 as an independent test data set (details in Methods).
        
        In order to create the \textit{positive} samples (time series containing an SSE), we modeled 30,000 dislocations (approximated as a point source) distributed along the Cascadia subduction interface (see Figure \ref{fig1}(b)) following the slab2 geometry \cite{hayes2018slab2} (detailed procedure in Methods). The focal mechanism of the synthetic ruptures approximates a thrust, with rake angle following a uniform distribution (from 75 to 100°) and strike and dip defined by the geometry of the slab. The magnitude of the synthetic SSEs is drawn from a uniform probability distribution (from $M_{w}$ 6 to 7). Their depths follow the slab geometry and are taken down to 60 km, with further variability of $\pm 10$ km. We further assign each event a realistic stress drop modeled from published scaling laws \cite{gao2012scaling}. We use the Okada dislocation model \cite{okada1985surface} to compute static displacements at each real GNSS station. We scaled the amplitudes of synthetic SSE signals, modeled as sigmoidal-shaped transients, with a duration following a uniform distribution (from 10 to 30 days). Eventually, we compute a database of 30,000 synthetic SSE transients, where the amplitude was added to the positive samples (placed in the middle of the 60-day window).
        
        The synthetic data set is thus made of 60,000 samples and labels, equally split into pure noise (labeled as 0) and signal (labeled as 1) with different nuances of signal-to-noise ratio, resulting both from different station noise levels and differences in magnitude and location, so that the deep learning method effectively learns to detect a variety of slow deformation transients from the background noise. The data set is further split into three independent training (60\%), validation (20\%) and test (20\%) sets, with the latter being used after the training phase only.
        
    \subsection*{SSEdetector: high-level architecture}
        
        SSEdetector is a deep neural network made of a CNN \cite{lecun2015deep}  and a Transformer network \cite{vaswani2017attention}  that are sequentially connected (detailed structure in Methods). We constructed the CNN to be a deep spatial-temporal encoder, that behaves as feature extractor. The structure of the encoder is a deep cascade of 1-dimensional temporal convolutional block sequences and spatial pooling layers. The depth of the feature extractor guarantees: (1) a high expressive power, \textit{i.e.}, detailed low-level spatiotemporal features, (2) robustness to data gaps, since their propagation is kept limited to the first layers thanks to a cascade of pooling operators, and (3) limited overfit of the model on the station patterns, thanks to the spatial pooling operation.
        The decisive component of our architecture is the Transformer network, placed right after the deep CNN encoder. The role of the Transformer is to apply a temporal self-attention mechanism to the features computed by the CNN. As humans, we instinctively focus just on particular fragments of data when looking for any specific patterns. We wanted to replicate such a behavior in our methodology, leading to a network able to enhance crucial portions of the data and neglect the irrelevant ones. This is done by assigning a weight to the data, with those weights being learnt from the data itself. As a result, our Transformer has learnt (1) to precisely identify the timing of the aseismic deformation transients in the geodetic time series and (2) to focus on it by assigning a weight close to zero to the rest of the time window.
        We further guide the process of finding slow deformation transients through a specific supervised-learning classification process. First, the disclosed outputs of the Transformer are averaged and passed through a sigmoid activation function. The output values are a detection probability lying in the (0,1) range and can be further interpreted as a confidence measure of the method. Second, we train SSEdetector by minimizing the binary cross-entropy loss between the target and the predicted labels (details in Methods). The combination of the two strategies allows SSEdetector to be successfully applied in a real context because: (1) we can run our detector on 1-day-shift windows of real data and collect an output value for each day used to build a temporal probability curve, (2) thanks to the Transformer neural network, such a curve will be smooth and the value of probability will gradually increase in time as SSEdetector identifies slow deformation in the geodetic data.
    
    \subsection*{Application to the synthetic test set: detection threshold}
        
        \begin{figure}
            \centering
            \includegraphics[width=1.\textwidth]{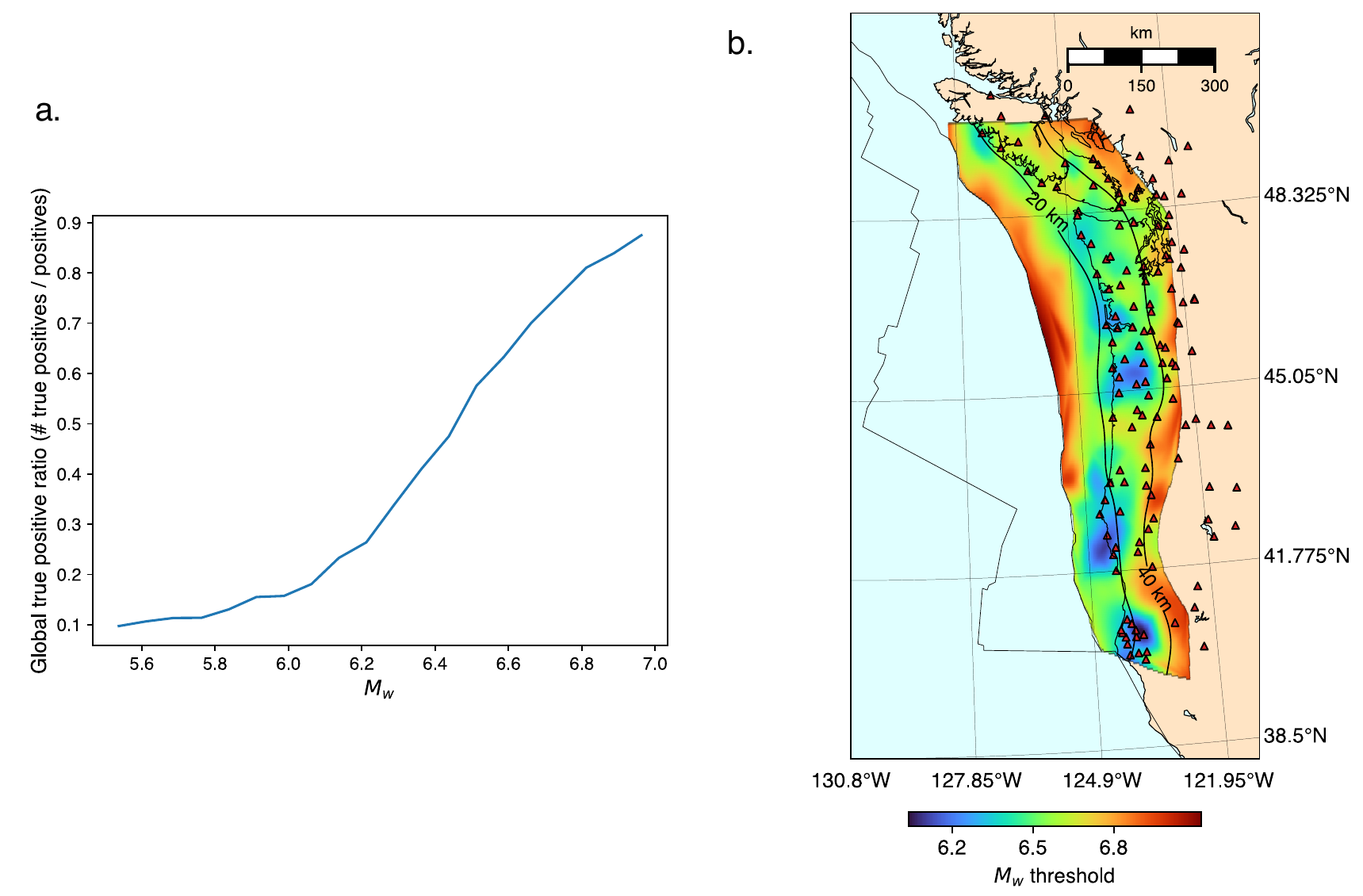}
            \caption{\textbf{Performance of SSEdetector on synthetic data}. (a) The blue curve represents the true positive rate (probability that an actual positive will test positive), computed on synthetic data, as a function of the magnitude. (b) Map showing the spatial distribution of the magnitude threshold for reliable detection, computed, for each spatial bin, as the minimum magnitude corresponding to a true positive rate value of 0.7.}\label{fig2}
        \end{figure}
        
        We test SSEdetector against unseen synthetic samples and we analyze the results quantitatively. We generate test synthetic samples from GNSS data in the period 2018-2022 to limit the influence of data gaps (details in Method). We obtain a measure of the sensitivity of our model by computing the true positive rate (TPR, probability that an actual positive will test positive) as a function of the magnitude. On a global scale, the sensitivity is increasing with the signal-to-noise ratio (SNR), which also shows that it exists an SNR threshold limit for any SSE detection. This threshold is mainly linked to the magnitude of the event, rather than the moment rate. Thus, the ability of SSE detection is mostly influenced by the signal-to-noise ratio rather than the event duration (cf. Supplementary Figure 1). We compute the sensitivity as a function of the spatial coordinates of the SSE, by deriving a synthetic proxy as the magnitude threshold under which the TPR is smaller than 0.7 on a spatial neighborhood of approximately 50 km. We can see from Figure \ref{fig2}(b) that the detection power is related to the density of stations in the GNSS network, as well as to the distance between the rupture and the nearest station, and the rupture depth. When the density of GNSS stations is not high enough, our resolution power decreases as well as the reliability of the prediction. In those cases, we can only detect high-magnitude SSEs. This is also the case on the eastern side of the targeted region where the SSE sources are deeper because of the slab geometry (Figure \ref{fig1}(b)), even in locations where the density of stations is higher. In this case, the magnitude threshold increases because these events are more difficult to detect.
        
    \subsection*{Continuous SSE detection in Cascadia from raw geodetic data during 2007-2022}
        \subsubsection*{Overall characteristics of the detected events}
            
            \begin{figure}
                \centering
                \includegraphics[width=1.\textwidth]{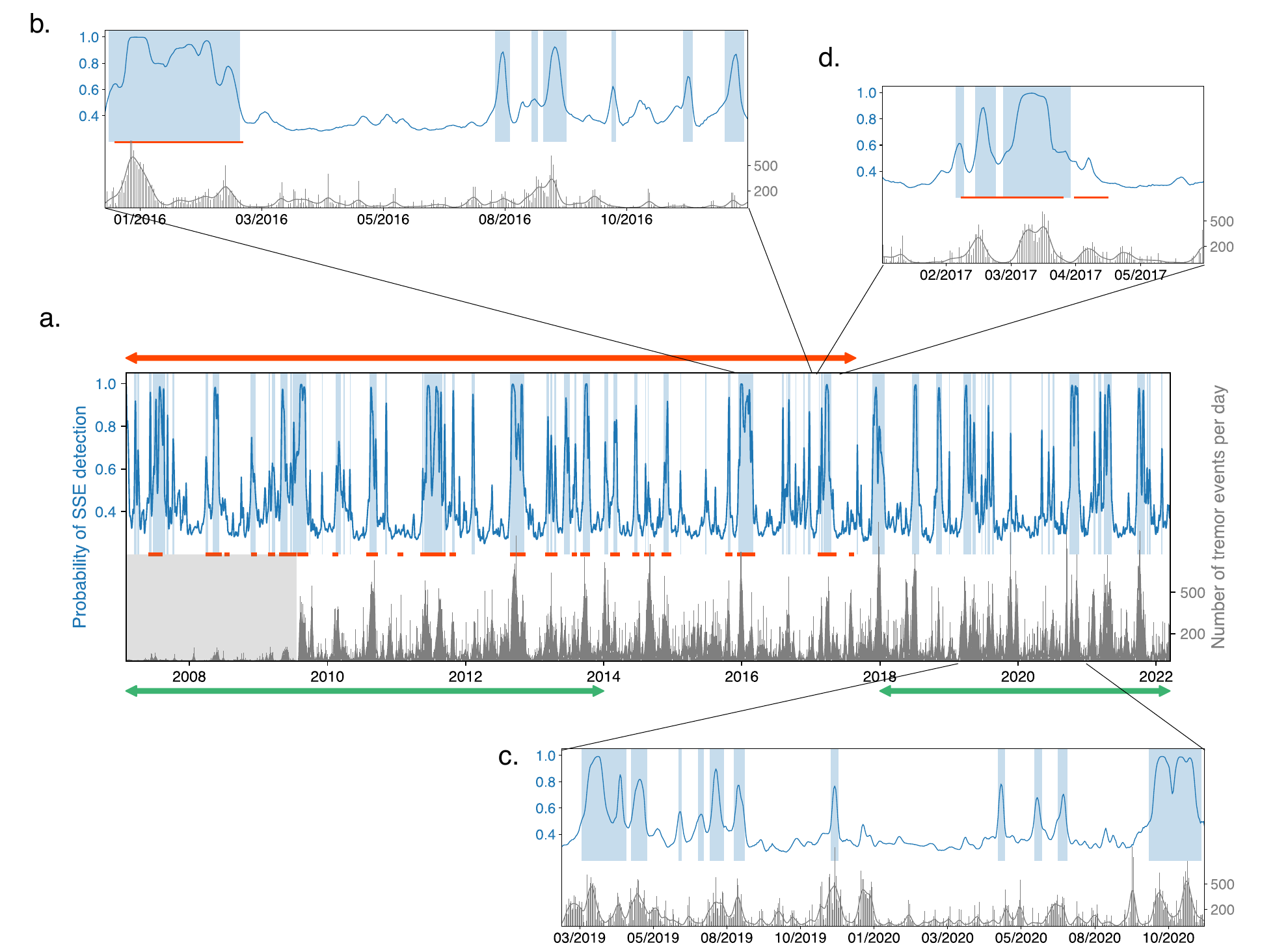}
                \caption{\textbf{Overview of the performance of SSEdetector on real raw GNSS time series}. The blue curves show the probability of detecting a slow slip event (output by SSEdetector) in 60-day sliding windows centered on a given date. Grey bars represent the number of tremors per day, smoothed (gaussian smoothing, $\sigma=2$ days) in the grey curve. Red horizontal segments represent the known events catalogued by \cite{Michel2019}. The (a) panel shows the global performance of SSEdetector over 2007-2022. The red arrow indicates the time window analyzed by Michel et al., while the green arrows describe the two periods from which the synthetic training samples have been derived. The grey rectangle indicates the period which was not covered by the PNSN catalog. In this period, data from Ide, 2012 \cite{ide2012variety} has been used. The (b), (c) and (d) panels show zooms on 2016-2017, 2019-2021 and 2017 (January to July), respectively.}\label{fig3}
            \end{figure}
            

            \begin{table}
                \caption{\textbf{Comparison of the number detections from SSEdetector with respect to the catalog from Michel et al. \cite{Michel2019}}. We distinguish detections in 2007-2017 (the same period analyzed by Michel et al.), and in 2017-2022. We further discriminate, in 2007-2017, between events in common with Michel et al., and new events in the same period.}
                \centering
                \begin{tabular}{cccc}
                    \multicolumn{2}{c}{Period} & \multicolumn{2}{c}{Method}\\
                    \hline
                    && Michel et al.  & SSEdetector \\
                    \hline
                    \multicolumn{2}{l}{\multirow{2}{*}{2007-2017}} Common with Michel et al. & 40 & 35\\
                    \multicolumn{2}{l}{} Not detected by Michel et al. & 0 & 20\\
                    \hline
                    2017-2022 & & 0 & 23 \\
                    \hline
                \label{tab:event-comparison-michel}
                \end{tabular}
            \end{table}

            In order to evaluate how SSEdetector performs on real continuous data, we applied it to the raw GNSS time series in Cascadia for the period 2007-2022. SSEdetector scans the data with a 60-day sliding window (1-day stride), providing a probability of detection for the central day in each window. Figure \ref{fig3}(a) shows the probability of slow slip event detection (in blue) together with the tremor activity over the period 2007-2022 (in grey). We consider having a reliable detection when the probability value exceeds 0.5. We find 78 slow slip events over the period 2007-2022, with durations ranging from 2 to 79 days. We find 55 slow slip events in the period 2007-2017, to be compared with the 40 detections of the catalog of Michel et al. \cite{Michel2019} (Table \ref{tab:event-comparison-michel}). We detect 35 of the 40 (87.5\%) catalogued SSEs. Three of the missed SSEs have a magnitude smaller than 5.5, one of them has a magnitude of 5.86. The remaining one has a magnitude of 6.03. We show their location in Supplementary Figure 2, superimposed on the magnitude threshold map derived for SSEdetector (see Figure \ref{fig2}(b)). Given their location, the five missed events have magnitudes that are below the magnitude resolution limit (from 6 to 6.5, see Supplementary Figure 2). The remaining 20 events may be associated with new undetected SSEs. We also find 23 new events in the period 2017-2022, which was not covered by Michel et al. \cite{Michel2019}. We fixed the detection threshold to its default value of 0.5, \textit{i.e.}, the model detects an event with a 50\% confidence. Yet, this threshold can be modified in accordance with specific needs: if high-confidence detections are required, the threshold can be raised; conversely, it can be lowered to capture more events with lower confidence. Interestingly, the few SSE from Michel et al. that were missed with a 0.5 confidence are all detected when selecting a 0.4 threshold.
            
            We also analyze the shape of the static displacement field in correspondence with the detected SSEs (cf. Supplementary Figure 3). We compute the static displacement field by taking the median displacement over three days and subtracting the displacement value at each station corresponding to the dates of the SSE. We find a good accordance with independent studies \cite{Itoh2022, Michel2019, bletery2020slip}. Moreover, many of the events found after 2018, as well as the new events detected in the period analyzed by Michel et al., have a displacement field suggesting that they are correct detections.
        
        \subsubsection*{Analysis of the SSE durations}
            
            \begin{figure}%
                \centering
                \includegraphics[width=1.\textwidth]{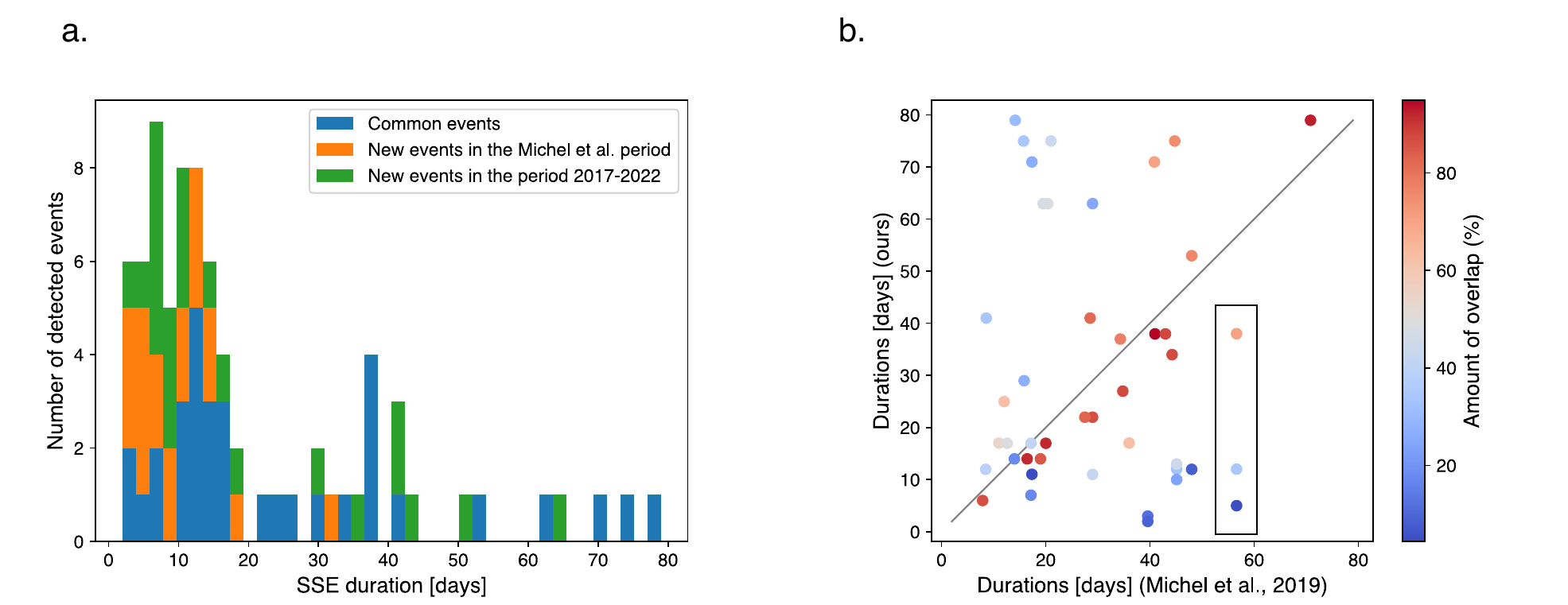}
                \caption{\textbf{Distribution of the detected SSEs and comparison with the independent catalogue from Michel et al. \cite{Michel2019}.} (a) Cumulative histogram of SSEdetector inferred durations. Blue bars represent the 35 catalogued events by Michel et al. \cite{Michel2019} that have been successfully retrieved. Orange bars show the 20 additional events that have been discovered within the time window analyzed by Michel et al., while green ones represent the 23 events found in the time period 2017-2022, not covered by the catalog of Michel et al. (b) Event durations from Michel et al.'s catalogue with respect to the durations obtained by SSEdetector. Events are color-coded by the overlap percentage (details in Methods).}\label{fig4}
            \end{figure}
            
            The shape of the probability curve gives insights into how SSEdetector reveals slow slip events from raw geodetic data. The probability curve in correspondence with an event has a bell shape: it grows until a maximum value, then it smoothly decreases when the model does not see any displacement associated with slow deformation in the data anymore. We use this property of the probability curve to extract a proxy on the detected SSE duration, based on the time span associated with the probability curve exceeding 0.5. We present the duration distribution in Figure \ref{fig4}. We detect most of the SSEs found by Michel et al., but we also find many more events, not only in the 2018-2022 period which was not investigated by Michel et al., but also within the 2007-2017 time window that they analyzed, suggesting that our method is more sensitive. We find potential slow slip events at all scales of durations (from 2 to 79 days). Michel et al. hardly detect SSEs that last less than 15 days, probably due to temporal data smoothing \cite{Michel2019}, while we retrieve shorter events (less than 10 days) since we use raw time series, meaning that our method has a better temporal resolution. In Figure \ref{fig4} (b), we show a comparison between the SSE durations of Michel et al.'s \cite{Michel2019} catalogued events and ours. This plot is made by considering all the combinations between events in our catalog and in the Michel et al. one. Each horizontal alignment represents an event in our catalogue that is split into sub-events in the Michel et al. catalog, while vertical alignments show events in the Michel et al. catalog corresponding to sub-events in our catalog. We find that the durations are in good accordance for a large number of events, for which the overlap is often higher than 70\%, both for small- and large-magnitude ones.
            We can also identify, from figure \ref{fig4}(b), that some events are separated in one method while identified as one single SSE in the other: this is the case for the $55$ day-long event from Michel et al. \cite{Michel2019}, that was paired with 3 SSEdetector sub-events (see Figure \ref{fig3}(d) and the rectangle in Figure \ref{fig4}(b)). The majority of the points located off the identity line (the diagonal) are thus sub-events for which the grouping differs in the two catalogs. As more points are below the diagonal than above, we can see that SSEdetector tends to separate the detections more. We interpret this as a possible increase in the detection precision, yet a validation with an independent acquisition data set is needed, since the separation into sub-events strongly depends on the threshold applied to the detection probability to define a slow slip event (0.5 in this study).
        
       \subsubsection*{Validation against tremors}
        
        \begin{figure}%
            \centering
            \includegraphics[width=1.\textwidth]{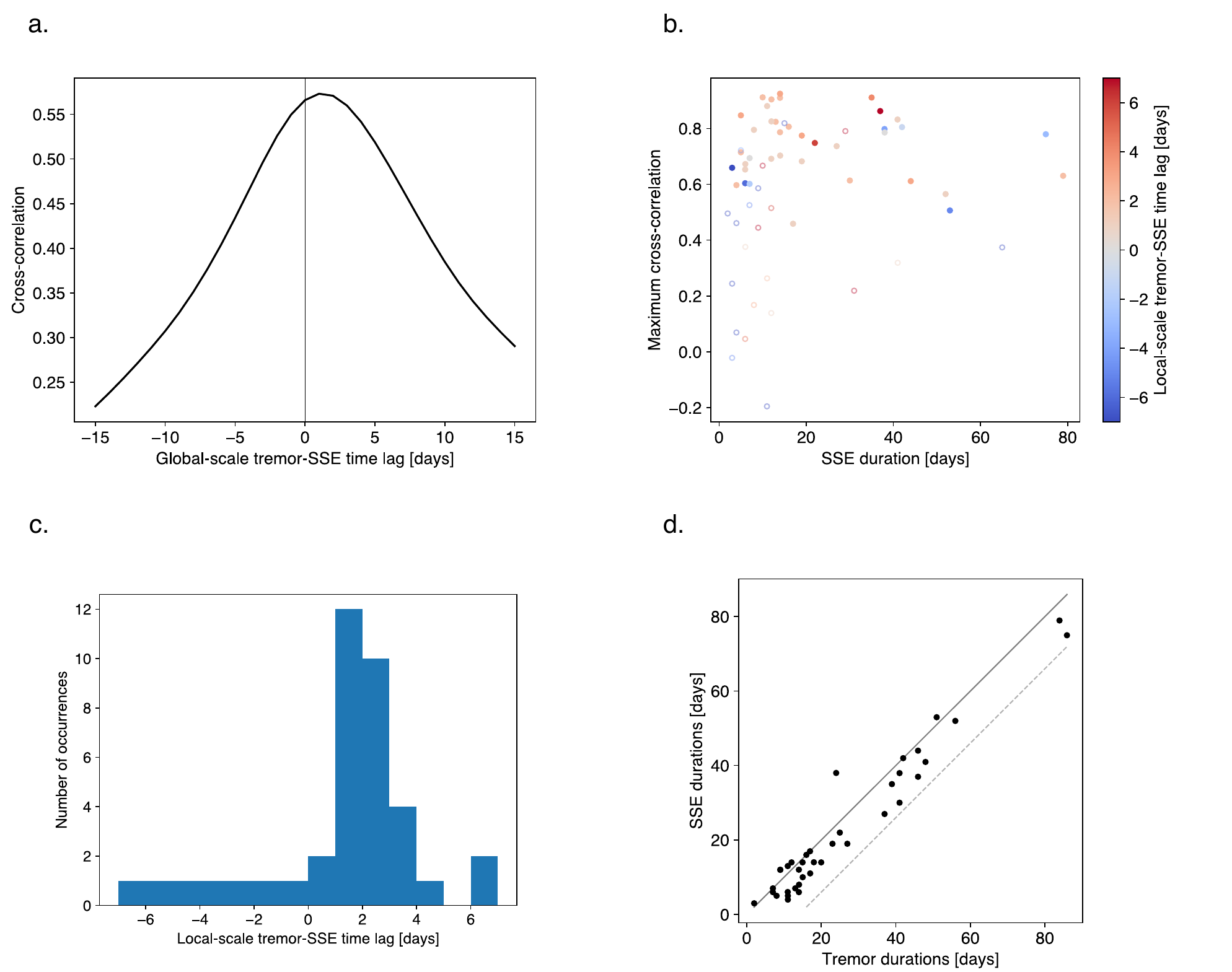}
            \caption{\textbf{Validation of SSEdetector performance against tremor activity in 2010-2022}. (a) Global-scale cross-correlation between the full-length SSEdetector output probability and the number of tremors per day, as a function of the time shift between the two curves. (b) Local-scale maximum value of cross-correlation for each SSE and tremor windows, centered on the SSE duration, as a function of the SSE duration, color-coded by the associated time lag between tremor and SSE (positive lag means deformation precedes tremor). Events having a zero-lag cross-correlation (correlation coefficient) lower than 0.4 are marked with an empty point. (c) Histogram of the time lags computed in the (b) panel. (d) SSE durations as a function of the tremor durations for the events in the (b) panel which have a correlation greater than 0.4. The solid black line represents the identity line, while the dashed grey line is the maximum tremor duration that can be attained for a given SSE duration, that is SSE duration + 14 days (see section section "\nameref{sec:local-global}").}\label{fig5}
        \end{figure}
        
        In order to have an independent validation, we compare our results with tremor activity from the Pacific Northwest Seismic Network (PNSN) catalog \cite{Wech2010} between 2009-2022 and Ide's catalogue \cite{ide2012variety} catalog between 2006-2009, shown in grey in Figure 
        \ref{fig3}. We show the location of the tremors in our catalogues with the dashed black contour in Figure \ref{fig1}(b). From a qualitative point of view, we can see that the detection probability curve seems to align well with the number of tremors per day, throughout the whole period. This is also true for the 20 possible new detected events that were not present in previous catalogs, for example during the period after 2017 (see Figure \ref{fig3}(c)), but also in 2016-2017, where we detect 11 possible events that were not previously catalogued (see Figure \ref{fig3}(b)). The excellent similarity between tremors and our detections is quantitatively assessed by computing the cross-correlation between the probability curve and the number of tremors per day, the latter smoothed with a gaussian filter ($\sigma = 1.5$ days), as a function of the time shift between the two curves (Figure \ref{fig5}(a)). The interval 2007-2010 has been excluded from Figure \ref{fig5}(a) in order to consider the period covered by the PNSN catalog only. The maximum correlation value is around 0.58 and is obtained for a time shift between 1 and 2 days. This shows that, at a global scale, the probability peaks are coeval with the peaks of tremor activity.
        
        We also make a further comparison at the local scale for each individual detected SSE. In Figure \ref{fig5}(b) we observe that most of the individual detected SSEs show a correlation larger than 0.4 with the coeval peak of tremor. SSE and tremor signals are offset by about 2 days on average (see Figure \ref{fig5}(c)).
        This result, obtained on windows of month-long scale, seems consistent with the decade-long correlation shown in Figure \ref{fig5}(a), suggesting that the found large-scale trend is also true at a smaller scale. This may suggest that the slow deformation, for which the detection probability is a proxy, precedes the tremor chatter by a few days, with potential implications on the nucleation of the slow rupture.
        
        We compare the tremor peak duration (see details in Methods) to the SSE duration in Figure \ref{fig5}(d) for all the events that have been also considered in Figure \ref{fig5}(b). The figure shows a correspondence between slow slip duration and coeval tremor activity duration: most of the events are associated with a peak of tremor activity of close duration. This is true also for large events, up to 80 days. This finding gives an insight that our deep learning-based method, blindly applied to raw geodetic time series, achieves reliable results. Yet, this result should be taken with caution, since it is strongly dependent on the choice of the window of observation (see Methods section for further details).

    \subsection*{Sensitivity study} \label{sec:sensitivity}
        
        We analyze the sensitivity of SSEdetector with respect to the number of stations. We construct an alternative test selecting 352 GNSS stations (see Supplementary Figure 5), which is the number of stations used by Michel et al. \cite{Michel2019}. The 217 extra stations have larger percentages of missing data compared to the initial 135 stations (cf. Supplementary Figure 4). We train and test SSEdetector with 352 time series and we report the results in Supplementary Figures 6-7. We observe that the results are similar, with an excellent alignment with tremors and similar correlation and lag values, although with this setting the detection power slightly decreases, probably due to a larger number of missing data. 
        
        We also test the ability of SSEdetector to identify SSEs in a sub-region only (even if it is trained with a large-scale network. For that, we test SSEdetector (trained on 135 stations), without re-training, on a subset of the GNSS network, situated in the northern part of Cascadia. To this end, we replace with zeros all the data associated with stations located at latitudes lower than $47$ degrees (see Supplementary Figure 8). Similarly, we find that SSEdetector retrieves all the events which were found by Michel et al. \cite{Michel2019} and the correlation with tremors that occur in this sub-region is still high, with a global-scale cross-correlation of 0.5 (cf. Supplementary Figures 9-10). This means that the model is robust against long periods of missing data and, thanks to the spatial pooling strategy, can generalize over different settings of stations and obtain some information on the localization.

        Finally, we test SSEdetector against other possible deep learning models that could be used for detection. We report in Supplementary Figures 11-12 the results obtained by replacing the one-dimensional convolutional layers with two-dimensional convolutions on time series sorted by latitude (as shown in Figure \ref{fig1}(a)). This type of architecture was used in studies having similar multi-station time-series data \cite{licciardi2022instantaneous}. We observe that the results on real data are not satisfactory because of too high a rate of false detections and a lower temporal resolution than SSEdetector (in other words, short SSEs are not retrieved). This suggests that our specific model architecture, handling in different ways the time dimension and the station dimension, might be more suited to multi-station time-series data sets.

    \subsection*{Discussion}

        In this study, we use a multi-station approach that proves efficient in detecting slow slip events in raw GNSS time series even in presence of SSE migrations \cite{bletery2020slip, Itoh2022, Michel2019}. Thanks to SSEdetector, we are able to detect 87 slow slip events with durations from 2 to 79 days, with an average limit magnitude of about 6.4 in north Cascadia and 6.2 in south Cascadia computed on the synthetic test set (see Figure \ref{fig2}(b)). The magnitude of the smallest detected SSE in common with Michel et al. is 5.42, with a corresponding duration of 8.5 days. One current limitation of this approach is that the location information is not directly inferred. In this direction, some efforts should be made in developing a method for characterizing slow slip events after the detection in order to have information on the location, but also on the magnitude, of the slow rupture.
        
        We apply our methodology to the Cascadia subduction zone because it is the area where independent benchmarks exist and it is thus possible to validate a new method. However, the applicability of SSEgenerator and SSEdetector to other subduction zones is possible. The current approach is, however, region-specific. In fact, the characteristics of the targeted zone affect the structure of the synthetic data, thus a method trained on a specific region could have poor performance if tested on another one without retraining. This problem can be addressed by generating multiple data sets associated with different regions and combining them for the training. Also, we focus on the Cascadia subduction zone, where not much regular seismicity occurs, making it a prototypical test zone when looking for slow earthquakes. When addressing other regions, such as Japan, for example, the influence of earthquakes or post-seismic relaxation signals could make the problem more complex. This extension goes beyond the scope of this study, yet we think that it will be essential to tackle this issue in order to use deep learning approaches for the detection of SSEs in any region.

\section*{Conclusions}
    
    We developed a powerful pipeline, composed of a realistic synthetic GNSS time-series generation, SSEgenerator, and a deep-learning classification model, SSEdetector, aimed to detect slow slip events from a series of raw GNSS time series measured by a station network. We built a new catalog of slow slip events in the Cascadia subduction zone by means of SSEdetector. We found 78 slow slip events from 2007 to 2022, 35 of which are in good accordance with the existing catalog \cite{Michel2019}. The detected SSEs have durations that range between a few days to a few months. The detection probability curve correlates well with the occurrence of tremor episodes, even in time periods where we found new events. The duration of our SSEs, for the 35 known events, as well as for the 43 new detections, are found to be similar to the coeval tremor duration. The comparison between tremors and SSEs also shows that, both at a local and a global temporal scale, the slow deformation may precede the tremor chatter by a few days, with potential implications on the link between a slow slip that could drive the rupture of nearby small seismic asperities. This is the first successful attempt to detect SSEs from raw GNSS time series, and we hope that this preliminary study will lead to the detection of SSEs in other active regions of the world.

\section*{Methods}

    \subsection*{SSEgenerator: data selection}
        We consider the 550 stations in the Cascadia subduction zone, belonging to the MAGNET GNSS network, and we select data from 2007 to 2022. We train SSEdetector with synthetic data whose source was affected by different noise and data gap patterns. We divide the data into three periods: 2007-2014, 2014-2018, and 2018-2022. In order to create a more diverse training set, data in the period 2007-2014 and 2018-2022 has been chosen as a source for synthetic data generation. The period 2014-2018 was left aside and used as an independent validation set for performance assessment on real data. Nonetheless, since synthetic data is performed by applying random transformations, a test on the whole sequence 2007-2022 is possible without overfitting.
        
        For the two periods 2007-2014 and 2018-2022, we sort the GNSS stations by the total number of missing data points and we choose the 135 stations affected by fewer data gaps as the final subset for our study. We make sure that stations having too high a noise do not appear in this subset. We select 135 stations since it represents a good compromise between the presence of data and the longest data gap sequence in a 60-day window. However, we also train and test SSEdetector on 352 stations (the same number used in the study by Michel et al. \cite{Michel2019}). We briefly discuss the results in the section "\nameref{sec:sensitivity}".
    
    \subsection*{SSEgenerator: Generation of ultra-realistic noise time series}
        Raw GNSS data is first detrended at each of the 135 stations, \textit{i.e.} the linear trend is removed, where the slope and the intercept are computed, for each station, without taking into account the data gaps, \textit{i.e.}, for each station the mean over time is calculated without considering the missing data points, and is removed from the series. A matrix containing all station time series $\mathbf{X} \in \mathbb{R}^{N_t \times N_s}$ is built, where $N_t$ is the temporal length of the input time series and $N_s$ is the number of stations. In this study, we use 2 components (N-S and E-W) and we apply the following procedure for each component independently. Each column of $\mathbf{X}$ contains a detrended time series. We proceed as follows. The $\mathbf{X}$ matrix is then re-projected in another vector space through a Principal Component Analysis (PCA), as follows. First, the data is centered. The mean vector is computed $\boldsymbol{\mu} \in \mathbb{R}^{N_s}$, such that $\mu_i$ is the mean of the i-th time series. The centered matrix is considered $\tilde{\mathbf{X}} = \mathbf{X} - \boldsymbol{\mu}$, and is decomposed through Singular Value Decomposition (SVD) to obtain the matrix of right singular vectors $\mathbf{V}$, which is the rotation matrix containing the spatial variability of the original vector space. We further rotate the data by means of this spatial matrix to obtain spatially-uncorrelated time series $\hat{\mathbf{X}} = \tilde{\mathbf{X}} \mathbf{V}$. Then, we produce $\hat{\mathbf{X}}_R$, a randomized version of $\hat{\mathbf{X}}$, by applying the iteratively-refined amplitude-adjusted Fourier transform (AAFT) method \cite{schreiber2000surrogate}, having globally the same power spectrum and amplitude distribution of the input data. The number of AAFT iterations has been experimentally set to 5. The surrogate time series are then back-projected in the original vector space to obtain $\mathbf{X}_R = \hat{\mathbf{X}}_R \mathbf{V}^T + \boldsymbol{\mu}$. We further enrich the randomized time series $\mathbf{X}_R$ by imprinting the real pattern of missing data for 70 \% of the synthetic data. We shuffle the data gaps before imprinting them to the data, such that SSEdetector can better generalize over unseen test data for the same station, which necessarily would have a different pattern of data gaps. We leave the remaining 30\% of the data as it is. We prefer not to use any interpolation method in order not to introduce new values in the data. Thus, we set all the missing data points to zero, which is a neutral value with respect to the trend of the data and the convolution operations performed by SSedetector.
        
        After this process, we generate sub-windows of noise time series as follows. Given the window length $W_L$, a uniformly distributed random variable is generated $s \sim \mathcal{U}(-W_L/2, W_L/2)$ and the data is circularly shifted by the amount $s$. Then, $\lfloor{N_t/W_L}\rfloor$ contiguous (non-overlapping) windows are obtained. The circular shift is needed in order for SSEdetector not to learn a fixed temporal pattern of data gaps. Finally, by knowing the desired number $N$ of noise windows to compute, the surrogate generation $\left\{\mathbf{X}_R\right\}_i$ can be repeated $\lceil{\frac{N}{\lfloor{N_t/W_L}\rfloor}}\rceil$ times. In our study, we generate $N=60,000$ synthetic samples, by calling the surrogate data generation 1,429 times and extracting 42 non-overlapping noise windows from each randomized time series.

    \subsection*{SSEgenerator: Modeling of synthetic slow slip events}
        We first generate synthetic displacements at all the 135 selected stations using Okada's dislocation model \cite{okada1985surface}. We draw random locations, strike and dip angles using the slab2 model \cite{hayes2018slab2} following the subduction geometry within the area of interest (see Figure \ref{fig1}(b)). We let the rake angle be a uniform random variable from 75 to 100 degrees, in order to have a variability around 90 degrees (thrust focal mechanism). For each (latitude, longitude) couple, we extract the corresponding depth from the slab and we add further variability, modeled as a uniformly distributed random variable from -10 to 10 km. We allow for this variability if the depth is at least 15 km, in order not to have ruptures that reach the surface. We associate each rupture with a magnitude $M_w$, uniformly generated in the range (6, 7), and we compute the equivalent moment as $M_0 = 10^{1.5 M_w + 9.1}$. As for the fault geometry, we rely on the circular crack approximation \cite{lay1995modern} to compute the fault radius as:
        
        \begin{equation}
            R = \left(\frac{7}{16} \frac{M_0}{\Delta\sigma}\right)^{1/3}
        \end{equation}
        
        where $\Delta\sigma$ is the static stress drop. We compute the average slip on the fault as:
        
        \begin{equation}
            \bar{u} = \frac{16}{7 \pi} \frac{\Delta\sigma}{\mu}
        \end{equation}
        
        where $\mu$ is the shear modulus. We assume $\mu=30$ GPa. By imposing that the surface of the crack must equal a rectangular dislocation of length $L$  and width $W$, we obtain $L = \sqrt{2\pi} R$. We assume that $W = L/2$. Finally, we model the stress drop as a lognormally-distributed random variable. We assume the average stress drop to be $\overline{\Delta\sigma} = 0.05$ MPa for the Cascadia subduction zone \cite{gao2012scaling}. We also assume that the coefficient of variation $c_V$, namely the ratio between the standard deviation and the mean, is equal to 10. Hence, we generate the static stress drop as $\Delta\sigma \sim \text{Lognormal}(\mu_N, \sigma_N^2)$, where $\mu_N$ and $\sigma_N$ are the mean and the standard deviation of the underlying normal distribution, respectively, that we derive as:
        
        \begin{equation}
            \sigma_N = \sqrt{\ln{(c_V^2 + 1)}}
        \end{equation}
        
        and
        
         \begin{equation}
            \mu_N = \ln{(\overline{\Delta\sigma})} - \sigma_N^2/2.
        \end{equation}
        
        We thus obtain the (horizontal) synthetic displacement vector $\mathbf{D}_s=(D_s^{N-S}, D_s^{E-W})$ at each station $s$. We model the temporal evolution of slow slip events as a logistic function. Let $D$ be the E-W displacement for simplicity. In this case, we model an SSE signal at a station $s$ as:
        
        \begin{equation}
            d_s(t) = \frac{D}{1 + e^{-\beta (t-t_0)}}
        \end{equation}
        
        where $\beta$ is a parameter associated with the growth rate of the curve and $t_0$ is the time corresponding to the inflection point of the logistic function. We assume $t_0=30$ days, so that the signal is centered in the 60-day window. We derive the parameter $\beta$ as a function of the slow slip event duration $T$. We can rewrite the duration as $T=t_{max} - t_{min}$, where $t_{max}$ is the time corresponding to the steady-state value of the signal (\textit{i.e.}, $D$), while $t_{min}$ is associated to the minimum (\textit{i.e.}, 0). Since these values are only asymptotically reached, we introduce a threshold $\gamma$, such that $t_{max}$ and $t_{min}$ are associated with $d_s(D-\gamma)$ and $d_s(\gamma)$, respectively. We choose $\gamma=0.01 \cdot D$. By rewriting the duration as $T=t_{max} - t_{min}$ and solving for $\beta$, we obtain:
        
        \begin{equation}
            \beta = \frac{2}{T} \ln{\left(\frac{D}{\gamma} - 1\right)}.
        \end{equation}
        
        Finally, we generate slow slip events having uniform duration $T$ between 10 and 30 days. We take half of the noise samples (30,000) and we create a positive sample (\textit{i.e.}, time series containing a slow slip event) as $\mathbf{X}_R + \mathbf{d}(t)$, where $\mathbf{d}(t)$ is a matrix containing all the modeled time series $d_s(t)$ for each station. We let $\mathbf{X}_R$ contain missing data. Therefore, we do not add the signal $d_s(t)$ where data should not be present.
        
    \subsection*{SSEdetector: Detailed architecture}
        SSEdetector is a deep neural network obtained by the combination of a convolutional and a Transformer neural network. The full architecture is shown in Supplementary Figure 13. The model takes input GNSS time series, which can be grouped as a matrix of shape $(N_s, N_t, N_c)$, where $N_s, N_t, N_c$ are the number of stations, window length and number of components, respectively. In this study, $N_s=135, N_t=60$ days and $N_c=2$ (N-S, E-W). The basic unit of this CNN is a Convolutional Block. It is made of a sequence of a one-dimensional convolutional layer in the temporal ($N_t$) dimension,  which computes $N_f$ feature maps by employing a $1 \times 5$ kernel, followed by a Batch Normalization \cite{ioffe2015batch} and a ReLu activation function \cite{agarap2018deep}. We will refer to this unit as ConvBlock($N_f$) for the rest of the paragraph (see Supplementary Figure 13). We alternate convolutional operations in the temporal dimension with pooling operations in the station dimension (max-pooling with a kernel of 3) and we replicate this structure as long as the spatial (station) dimension is reduced to 1. To this end, we create a sequence of 3 ConvBlock($\cdot$) + max-pooling. As an example, the number of stations after the first pooling layer is reduced from 135 to 45. At each ConvBlock($\cdot$), we multiply by 4 the number of computed feature maps $N_{f}$. At the end of the CNN, the computed features have shape ($N_t, N_f^{final}$), with $N_f^{final}=256$. 
        
        This feature matrix is given as input to a Transformer neural network. We first use a Positional Embedding to encode the temporal sequence. We do not impose any kind of pre-computed embedding, but we use a learnable matrix of shape ($N_t, N_f^{final}$). The learnt embeddings are added to the feature matrix (\textit{i.e.}, the output of the CNN). The embedded inputs are then fed to a Transformer neural network \cite{vaswani2017attention}, whose architecture is detailed in Supplementary Figure 14. Here, the global (additive) self-attention of the embedded CNN features is computed as:
        
        \begin{equation}
            \eta_{t_1,t_2} = W_a \tanh{\left(W_1^T h_{t_1} + W_2^T h_{t_2} + b_h\right)} + b_a \ ,
        \end{equation}
        
        \begin{equation}
            a_{t_1,t_2} = \text{softmax}\left(\eta_{t_1,t_2}\right) = \frac{e^{\eta_{t_1,t_2}}}{\sum_{t_2}{e^{\eta_{t_1,t_2}}}} \ ,
        \end{equation}
        
        \begin{equation}
            c_{t_1} = \sum_{t_2}{a_{t_1,t_2} \cdot h_{t_2}} \ ,
        \end{equation}
        
        where $W$ represents a learnable weight matrix and $b$ a bias vector. The matrices $h_{t_1}$ and $h_{t_2}$ are the hidden-state representations at time $t_1$ and $t_2$, respectively. The matrix $a_{t_1,t_2}$ contains the attention scores for the time steps $t_1$ and $t_2$. Here, a context vector is computed as the weighted sum of the hidden-state representations by the attention scores. The context vector contains the importance at a given time step based on all the features in the window. The contextual information is then added to the Transformer inputs. Then, a position-wise Feed-Forward layer (with a dropout rate of $0.1$) is employed to add further non-linearity. After the Transformer network, a Global Average Pooling in the temporal dimension ($N_t$) is employed to gather the transformed features and to output a vector summarizing the temporal information. A Dropout is then added as a form of regularization to reduce overfitting \cite{srivastava2014dropout}, with dropout rate $\delta=0.2$. In the end, we use a fully-connected layer with one output, with a sigmoid activation function to express the probability of SSE detection.

    \subsection*{Training details}

        We perform a mini-batch training \cite{bottou2018optimization} (batch size of 128 samples) by minimizing the binary cross-entropy (BCE) loss between the target labels $y$ and the predictions $\hat{y}$ (a probability estimate):
        
        \begin{equation}
            \text{BCE}(y, \hat{y}) = - y \ln(\hat{y}) - (1-y) \ln(1- \hat{y}).
        \end{equation}
        
        The BCE loss is commonly used for binary classification problems (detection is a binary classification). We use the ADAM method for the optimization \cite{kingma2014adam} with a learning rate $\lambda = 10^{-3}$ which has been experimentally chosen. We schedule the learning rate such that it is reduced during training iterations and we stop the training when the validation loss did not improve for 50 consecutive epochs. We initialized the weights of SSEdetector with a uniform He initializer \cite{he2015delving}. We implemented the code of SSEdetector in Python using the Tensorflow and Keras libraries \cite{chollet2015keras, abadi2016tensorflow}. We run the training on NVIDIA Tesla A100 Graphics Processing Units (GPUs). The training of SSEdetector takes less than 2 hours. The inference on the whole 15-year sequence (2007-2022) takes a few minutes.
        
    \subsection*{Calculation of tremor durations}\label{sec:tremor-dur}
        We compute the durations of tremor bursts using the notion of topographic prominence, explained in the following. We rely on the software implementation from the SciPy Python library \cite{2020SciPy-NMeth}. Given a peak in the curve, the topographic prominence is informally defined as the minimum elevation that needs to be descended to start reaching a higher peak. The procedure is graphically detailed in Supplementary Figure 15. We first search for peaks in the number of tremors per day by comparison with neighboring values. In order to avoid too many spurious local maxima, we smooth the number of tremors per day with a gaussian filter ($\sigma=1.5$ days). For each detected SSE, we search for peaks of tremors in a window given by the SSE duration $\pm 3$ days. For each peak of tremors that is found, the corresponding width is computed as follows. The topographic prominence is computed by placing a horizontal line at the peak height $h$ (the value of the tremor curve corresponding to the peak). An interval is defined, corresponding to the points where the line crosses either the signal bounds or the signal at the slope towards a higher peak. In this interval, the minimum values of the signal on each side are computed, representing the bases of the peak. The topographic prominence $p$ of the peak is then defined as the height between the peak and its highest base value. Then, the local height of the peak is computed as $h_L = h - \alpha \cdot p $. We set $\alpha=0.7$ in order to focus on the main tremor pulses, discarding further noise in the curve. From the local height, another horizontal line is considered and the peak width is computed as the intersection point of the line with either a slope, the vertical position of the bases or the signal bounds, on both sides. Finally, the total width of a tremor pulse in an SSE window is computed by considering the earliest starting point on the left side and the latest ending point on the right side.
        It must be noticed that, the derivation of the tremor duration depends on the window length. In fact, the inferred tremor duration can saturate to a maximum value equal to the length of the window. For this reason, we added in Figure \ref{fig5}(c) a dashed line corresponding to the window length (SSE duration + 14 days) (see section section "\nameref{sec:local-global}").

    \subsection*{Computation of local- and global-scale correlations} \label{sec:local-global}
        
        
        We compute the time-lagged cross-correlation between the SSE probability and the number of tremors per day (Fig. \ref{fig5}(a) and (b)). We smooth the number of tremors per day with a gaussian filter ($\sigma=1.5$ days). We consider a lag between -7 and 7 days, with a 1-day stride.
        
        In the case of Fig. \ref{fig5}(a), we compute the global correlation coefficient by considering the whole time sequence (2010-2022). As for Fig. \ref{fig5}(b), we make a local analysis. For each detected SSE, we first extract SSE and tremor slices from intervals centered on the SSE dates $[t_{SSE}^{start} - \Delta t, t_{SSE}^{end} + \Delta t]$, where $\Delta t = 30$ days. We first compute the cross-correlation between the two curves to filter out detected SSEs whose similarity with tremors is not statistically significant. For each SSE date, we discard $p''(t)$ and $n_T''(t)$ if their correlation coefficient is lower than 0.4. We build Fig. \ref{fig5}(b) after this process.

        We compute Fig. \ref{fig5}(d) by comparing the SSE and tremor durations for all the events that had a cross-correlation higher than 0.4. For those, we infer the tremor duration, using the method explained in section "\nameref{sec:tremor-dur}" on the $n_T(t)$ cut from an interval $[t_{SSE}^{start} - \Delta t', t_{SSE}^{end} + \Delta t']$, with $\Delta t' = 7$ days.

    \subsection*{Overlap percentage calculation}
    In Figure \ref{fig4} (b) we color-code the SSE durations by the overlap percentage between a pair of events, which we compute as the difference between the earliest end and the latest start, divided by the sum of the event lengths. Let $E_1$ and $E_2$ be two events with start and end dates given by $(t_1^{start}, t_1^{end})$ and $(t_2^{start}, t_2^{end})$ and with durations given by $D_1 = t_1^{end} - t_1^{start}$ and $D_2 = t_2^{end} - t_2^{start}$, respectively. We compute their overlap $\pi$ as:

    \begin{equation}
        \pi = \frac{\max(0, \min (t_1^{end}, t_2^{end}) - \max(t_1^{start}, t_2^{start}))}{D_1 + D_2}
    \end{equation}

\backmatter

\bmhead{Supplementary information}

This article has an accompanying supplementary file.

\bmhead{Acknowledgments}

    This work has been funded by ERC CoG 865963 DEEP-trigger. Most of the computations presented in this paper were performed using the GRICAD infrastructure (https://gricad.univ-grenoble-alpes.fr), which is supported by Grenoble research communities.

\section*{Declarations}


\begin{itemize}
\item \textbf{Funding}
    This work has been funded by ERC CoG 865963 DEEP-trigger.
\item \textbf{Competing interests}
    The authors declare no competing interests.
\item \textbf{Materials \& Correspondence}
    Correspondence to: Giuseppe Costantino
\item \textbf{Data availability}
    We downloaded the data from the Nevada Geodetic Laboratory (http://geodesy.unr.edu).
\item \textbf{Code availability}
    The source code of SSEgenerator and SSEdetector as well as the pre-trained model of SSEdetector are available at https://gricad-gitlab.univ-grenoble-alpes.fr/costangi/sse-detection.
\item \textbf{Authors' contributions}
    G.C. developed SSEgenerator and SSEdetector and produced all the results and figures presented here. A.S. designed the study and provided expertise for the geodetic data analysis. S.G.R. provided expertise for the Deep Learning aspects. G.C. wrote the first draft of the paper. All the authors contributed to reviewing the manuscript.
\end{itemize}

\noindent

\bibliography{sn-bibliography}


\begin{thebibliography}{44}
\ifx \bisbn   \undefined \def \bisbn  #1{ISBN #1}\fi
\ifx \binits  \undefined \def \binits#1{#1}\fi
\ifx \bauthor  \undefined \def \bauthor#1{#1}\fi
\ifx \batitle  \undefined \def \batitle#1{#1}\fi
\ifx \bjtitle  \undefined \def \bjtitle#1{#1}\fi
\ifx \bvolume  \undefined \def \bvolume#1{\textbf{#1}}\fi
\ifx \byear  \undefined \def \byear#1{#1}\fi
\ifx \bissue  \undefined \def \bissue#1{#1}\fi
\ifx \bfpage  \undefined \def \bfpage#1{#1}\fi
\ifx \blpage  \undefined \def \blpage #1{#1}\fi
\ifx \burl  \undefined \def \burl#1{\textsf{#1}}\fi
\ifx \doiurl  \undefined \def \doiurl#1{\url{https://doi.org/#1}}\fi
\ifx \betal  \undefined \def \betal{\textit{et al.}}\fi
\ifx \binstitute  \undefined \def \binstitute#1{#1}\fi
\ifx \binstitutionaled  \undefined \def \binstitutionaled#1{#1}\fi
\ifx \bctitle  \undefined \def \bctitle#1{#1}\fi
\ifx \beditor  \undefined \def \beditor#1{#1}\fi
\ifx \bpublisher  \undefined \def \bpublisher#1{#1}\fi
\ifx \bbtitle  \undefined \def \bbtitle#1{#1}\fi
\ifx \bedition  \undefined \def \bedition#1{#1}\fi
\ifx \bseriesno  \undefined \def \bseriesno#1{#1}\fi
\ifx \blocation  \undefined \def \blocation#1{#1}\fi
\ifx \bsertitle  \undefined \def \bsertitle#1{#1}\fi
\ifx \bsnm \undefined \def \bsnm#1{#1}\fi
\ifx \bsuffix \undefined \def \bsuffix#1{#1}\fi
\ifx \bparticle \undefined \def \bparticle#1{#1}\fi
\ifx \barticle \undefined \def \barticle#1{#1}\fi
\bibcommenthead
\ifx \bconfdate \undefined \def \bconfdate #1{#1}\fi
\ifx \botherref \undefined \def \botherref #1{#1}\fi
\ifx \url \undefined \def \url#1{\textsf{#1}}\fi
\ifx \bchapter \undefined \def \bchapter#1{#1}\fi
\ifx \bbook \undefined \def \bbook#1{#1}\fi
\ifx \bcomment \undefined \def \bcomment#1{#1}\fi
\ifx \oauthor \undefined \def \oauthor#1{#1}\fi
\ifx \citeauthoryear \undefined \def \citeauthoryear#1{#1}\fi
\ifx \endbibitem  \undefined \def \endbibitem {}\fi
\ifx \bconflocation  \undefined \def \bconflocation#1{#1}\fi
\ifx \arxivurl  \undefined \def \arxivurl#1{\textsf{#1}}\fi
\csname PreBibitemsHook\endcsname

\bibitem{Dragert2001}
\begin{barticle}
\bauthor{\bsnm{Dragert}, \binits{H.}},
\bauthor{\bsnm{Wang}, \binits{K.}},
\bauthor{\bsnm{James}, \binits{T.S.}}:
\batitle{A silent slip event on the deeper cascadia subduction interface}.
\bjtitle{Science}
\bvolume{292},
\bfpage{1525}--\blpage{1528}
(\byear{2001}).
\doiurl{10.1126/SCIENCE.1060152/ASSET/DD167B7B-24D4-40A4-996F-4603D42C0244/ASSETS/GRAPHIC/SE1919443004.JPEG}
\end{barticle}
\endbibitem

\bibitem{Lowry2001}
\begin{barticle}
\bauthor{\bsnm{Lowry}, \binits{A.R.}},
\bauthor{\bsnm{Larson}, \binits{K.M.}},
\bauthor{\bsnm{Kostoglodov}, \binits{V.}},
\bauthor{\bsnm{Bilham}, \binits{R.}}:
\batitle{Transient fault slip in guerrero, southern mexico}.
\bjtitle{Geophysical Research Letters}
\bvolume{28},
\bfpage{3753}--\blpage{3756}
(\byear{2001}).
\doiurl{10.1029/2001GL013238}
\end{barticle}
\endbibitem

\bibitem{Schwartz2007}
\begin{botherref}
\oauthor{\bsnm{Schwartz}, \binits{S.Y.}},
\oauthor{\bsnm{Rokosky}, \binits{J.M.}}:
Slow slip events and seismic tremor at circum-pacific subduction zones.
Reviews of Geophysics
\textbf{45}
(2007).
\doiurl{10.1029/2006RG000208}
\end{botherref}
\endbibitem

\bibitem{Ide2007}
\begin{barticle}
\bauthor{\bsnm{Ide}, \binits{S.}},
\bauthor{\bsnm{Beroza}, \binits{G.C.}},
\bauthor{\bsnm{Shelly}, \binits{D.R.}},
\bauthor{\bsnm{Uchide}, \binits{T.}}:
\batitle{A scaling law for slow earthquakes}.
\bjtitle{Nature 2007 447:7140}
\bvolume{447},
\bfpage{76}--\blpage{79}
(\byear{2007}).
\doiurl{10.1038/nature05780}
\end{barticle}
\endbibitem

\bibitem{Mousavi2020}
\begin{botherref}
\oauthor{\bsnm{Mousavi}, \binits{S.M.}},
\oauthor{\bsnm{Ellsworth}, \binits{W.L.}},
\oauthor{\bsnm{Zhu}, \binits{W.}},
\oauthor{\bsnm{Chuang}, \binits{L.Y.}},
\oauthor{\bsnm{Beroza}, \binits{G.C.}}:
Earthquake transformer—an attentive deep-learning model for simultaneous
  earthquake detection and phase picking.
Nature Communications
\textbf{11}
(2020).
\doiurl{10.1038/s41467-020-17591-w}
\end{botherref}
\endbibitem

\bibitem{Rousset2017}
\begin{barticle}
\bauthor{\bsnm{Rousset}, \binits{B.}},
\bauthor{\bsnm{Campillo}, \binits{M.}},
\bauthor{\bsnm{Lasserre}, \binits{C.}},
\bauthor{\bsnm{Frank}, \binits{W.B.}},
\bauthor{\bsnm{Cotte}, \binits{N.}},
\bauthor{\bsnm{Walpersdorf}, \binits{A.}},
\bauthor{\bsnm{Socquet}, \binits{A.}},
\bauthor{\bsnm{Kostoglodov}, \binits{V.}}:
\batitle{A geodetic matched filter search for slow slip with application to the
  mexico subduction zone}.
\bjtitle{Journal of Geophysical Research: Solid Earth}
\bvolume{122},
\bfpage{10498}--\blpage{10514}
(\byear{2017}).
\doiurl{10.1002/2017JB014448}
\end{barticle}
\endbibitem

\bibitem{Frank2015}
\begin{barticle}
\bauthor{\bsnm{Frank}, \binits{W.B.}},
\bauthor{\bsnm{Radiguet}, \binits{M.}},
\bauthor{\bsnm{Rousset}, \binits{B.}},
\bauthor{\bsnm{Shapiro}, \binits{N.M.}},
\bauthor{\bsnm{Husker}, \binits{A.L.}},
\bauthor{\bsnm{Kostoglodov}, \binits{V.}},
\bauthor{\bsnm{Cotte}, \binits{N.}},
\bauthor{\bsnm{Campillo}, \binits{M.}}:
\batitle{Uncovering the geodetic signature of silent slip through repeating
  earthquakes}.
\bjtitle{Geophysical Research Letters}
\bvolume{42},
\bfpage{2774}--\blpage{2779}
(\byear{2015}).
\doiurl{10.1002/2015GL063685}
\end{barticle}
\endbibitem

\bibitem{Gomberg2016}
\begin{barticle}
\bauthor{\bsnm{Gomberg}, \binits{J.}},
\bauthor{\bsnm{Wech}, \binits{A.}},
\bauthor{\bsnm{Creager}, \binits{K.}},
\bauthor{\bsnm{Obara}, \binits{K.}},
\bauthor{\bsnm{Agnew}, \binits{D.}}:
\batitle{Reconsidering earthquake scaling}.
\bjtitle{Geophysical Research Letters}
\bvolume{43},
\bfpage{6243}--\blpage{6251}
(\byear{2016}).
\doiurl{10.1002/2016GL069967}
\end{barticle}
\endbibitem

\bibitem{Hawthorne2018}
\begin{barticle}
\bauthor{\bsnm{Hawthorne}, \binits{J.C.}},
\bauthor{\bsnm{Bartlow}, \binits{N.M.}}:
\batitle{Observing and modeling the spectrum of a slow slip event}.
\bjtitle{Journal of Geophysical Research: Solid Earth}
\bvolume{123},
\bfpage{4243}--\blpage{4265}
(\byear{2018}).
\doiurl{10.1029/2017JB015124}
\end{barticle}
\endbibitem

\bibitem{Frank2019}
\begin{botherref}
\oauthor{\bsnm{Frank}, \binits{W.B.}},
\oauthor{\bsnm{Brodsky}, \binits{E.E.}}:
Daily measurement of slow slip from low-frequency earthquakes is consistent
  with ordinary earthquake scaling.
Science Advances
\textbf{5}
(2019).
\doiurl{10.1126/SCIADV.AAW9386/SUPPL_FILE/AAW9386_SM.PDF}
\end{botherref}
\endbibitem

\bibitem{Michel2019}
\begin{barticle}
\bauthor{\bsnm{Michel}, \binits{S.}},
\bauthor{\bsnm{Gualandi}, \binits{A.}},
\bauthor{\bsnm{Avouac}, \binits{J.P.}}:
\batitle{Similar scaling laws for earthquakes and cascadia slow-slip events}.
\bjtitle{Nature}
\bvolume{574},
\bfpage{522}--\blpage{526}
(\byear{2019}).
\doiurl{10.1038/s41586-019-1673-6}
\end{barticle}
\endbibitem

\bibitem{Bartlow2011}
\begin{botherref}
\oauthor{\bsnm{Bartlow}, \binits{N.M.}},
\oauthor{\bsnm{Miyazaki}, \binits{S.}},
\oauthor{\bsnm{Bradley}, \binits{A.M.}},
\oauthor{\bsnm{Segall}, \binits{P.}}:
Space-time correlation of slip and tremor during the 2009 cascadia slow slip
  event.
Geophysical Research Letters
\textbf{38}
(2011).
\doiurl{10.1029/2011GL048714}
\end{botherref}
\endbibitem

\bibitem{Radiguet2012}
\begin{barticle}
\bauthor{\bsnm{Radiguet}, \binits{M.}},
\bauthor{\bsnm{Cotton}, \binits{F.}},
\bauthor{\bsnm{Vergnolle}, \binits{M.}},
\bauthor{\bsnm{Campillo}, \binits{M.}},
\bauthor{\bsnm{Walpersdorf}, \binits{A.}},
\bauthor{\bsnm{Cotte}, \binits{N.}},
\bauthor{\bsnm{Kostoglodov}, \binits{V.}}:
\batitle{Slow slip events and strain accumulation in the guerrero gap, mexico}.
\bjtitle{Journal of Geophysical Research: Solid Earth}
\bvolume{117},
\bfpage{4305}
(\byear{2012}).
\doiurl{10.1029/2011JB008801}
\end{barticle}
\endbibitem

\bibitem{Kong2019}
\begin{barticle}
\bauthor{\bsnm{Kong}, \binits{Q.}},
\bauthor{\bsnm{Trugman}, \binits{D.T.}},
\bauthor{\bsnm{Ross}, \binits{Z.E.}},
\bauthor{\bsnm{Bianco}, \binits{M.J.}},
\bauthor{\bsnm{Meade}, \binits{B.J.}},
\bauthor{\bsnm{Gerstoft}, \binits{P.}}:
\batitle{Machine learning in seismology: Turning data into insights}.
\bjtitle{Seismological Research Letters}
\bvolume{90},
\bfpage{3}--\blpage{14}
(\byear{2019}).
\doiurl{10.1785/0220180259}
\end{barticle}
\endbibitem

\bibitem{Zhu2019}
\begin{barticle}
\bauthor{\bsnm{Zhu}, \binits{W.}},
\bauthor{\bsnm{Beroza}, \binits{G.C.}}:
\batitle{Phasenet: A deep-neural-network-based seismic arrival-time picking
  method}.
\bjtitle{Geophysical Journal International}
\bvolume{216},
\bfpage{261}--\blpage{273}
(\byear{2019}).
\doiurl{10.1093/gji/ggy423}
\end{barticle}
\endbibitem

\bibitem{Woollam2022}
\begin{barticle}
\bauthor{\bsnm{Woollam}, \binits{J.}},
\bauthor{\bsnm{Münchmeyer}, \binits{J.}},
\bauthor{\bsnm{Tilmann}, \binits{F.}},
\bauthor{\bsnm{Rietbrock}, \binits{A.}},
\bauthor{\bsnm{Lange}, \binits{D.}},
\bauthor{\bsnm{Bornstein}, \binits{T.}},
\bauthor{\bsnm{Diehl}, \binits{T.}},
\bauthor{\bsnm{Giunchi}, \binits{C.}},
\bauthor{\bsnm{Haslinger}, \binits{F.}},
\bauthor{\bsnm{Jozinović}, \binits{D.}},
\bauthor{\bsnm{Michelini}, \binits{A.}},
\bauthor{\bsnm{Saul}, \binits{J.}},
\bauthor{\bsnm{Soto}, \binits{H.}}:
\batitle{Seisbench-a toolbox for machine learning in seismology}.
\bjtitle{Seismological Research Letters}
\bvolume{93},
\bfpage{1695}--\blpage{1709}
(\byear{2022}).
\doiurl{10.1785/0220210324}
\end{barticle}
\endbibitem

\bibitem{Ross2019}
\begin{barticle}
\bauthor{\bsnm{Ross}, \binits{Z.E.}},
\bauthor{\bsnm{Trugman}, \binits{D.T.}},
\bauthor{\bsnm{Hauksson}, \binits{E.}},
\bauthor{\bsnm{Shearer}, \binits{P.M.}}:
\batitle{Searching for hidden earthquakes in southern california}.
\bjtitle{Science}
(\byear{2019}).
\doiurl{10.1126/SCIENCE.AAW6888/SUPPL_FILE/AAW6888_ROSS_SM.PDF}
\end{barticle}
\endbibitem

\bibitem{Tan2021}
\begin{barticle}
\bauthor{\bsnm{Tan}, \binits{Y.J.}},
\bauthor{\bsnm{Waldhauser}, \binits{F.}},
\bauthor{\bsnm{Ellsworth}, \binits{W.L.}},
\bauthor{\bsnm{Zhang}, \binits{M.}},
\bauthor{\bsnm{Zhu}, \binits{W.}},
\bauthor{\bsnm{Michele}, \binits{M.}},
\bauthor{\bsnm{Chiaraluce}, \binits{L.}},
\bauthor{\bsnm{Beroza}, \binits{G.C.}},
\bauthor{\bsnm{Segou}, \binits{M.}}:
\batitle{Machine‐learning‐based high‐resolution earthquake catalog
  reveals how complex fault structures were activated during the 2016–2017
  central italy sequence}.
\bjtitle{The Seismic Record}
\bvolume{1},
\bfpage{11}--\blpage{19}
(\byear{2021}).
\doiurl{10.1785/0320210001}
\end{barticle}
\endbibitem

\bibitem{Ross2020}
\begin{barticle}
\bauthor{\bsnm{Ross}, \binits{Z.E.}},
\bauthor{\bsnm{Cochran}, \binits{E.S.}},
\bauthor{\bsnm{Trugman}, \binits{D.T.}},
\bauthor{\bsnm{Smith}, \binits{J.D.}}:
\batitle{3d fault architecture controls the dynamism of earthquake swarms}.
\bjtitle{Science}
\bvolume{368},
\bfpage{1357}--\blpage{1361}
(\byear{2020}).
\doiurl{10.1126/SCIENCE.ABB0779/SUPPL_FILE/ABB0779_ROSS_SM.PDF}
\end{barticle}
\endbibitem

\bibitem{Tan2020}
\begin{barticle}
\bauthor{\bsnm{Tan}, \binits{Y.J.}},
\bauthor{\bsnm{Marsan}, \binits{D.}}:
\batitle{Connecting a broad spectrum of transient slip on the san andreas
  fault}.
\bjtitle{Science Advances}
\bvolume{6},
\bfpage{2489}--\blpage{2503}
(\byear{2020}).
\doiurl{10.1126/SCIADV.ABB2489/SUPPL_FILE/ABB2489_SM.PDF}
\end{barticle}
\endbibitem

\bibitem{rouet2021}
\begin{barticle}
\bauthor{\bsnm{Rouet-Leduc}, \binits{B.}},
\bauthor{\bsnm{Jolivet}, \binits{R.}},
\bauthor{\bsnm{Dalaison}, \binits{M.}},
\bauthor{\bsnm{Johnson}, \binits{P.A.}},
\bauthor{\bsnm{Hulbert}, \binits{C.}}:
\batitle{Autonomous extraction of millimeter-scale deformation in insar time
  series using deep learning}.
\bjtitle{Nature Communications 2021 12:1}
\bvolume{12},
\bfpage{1}--\blpage{11}
(\byear{2021}).
\doiurl{10.1038/s41467-021-26254-3}
\end{barticle}
\endbibitem

\bibitem{costantino2022seismic}
\begin{botherref}
\oauthor{\bsnm{Costantino}, \binits{G.}},
\oauthor{\bsnm{Giffard-Roisin}, \binits{S.}},
\oauthor{\bsnm{Marsan}, \binits{D.}},
\oauthor{\bsnm{Marill}, \binits{L.}},
\oauthor{\bsnm{Radiguet}, \binits{M.}},
\oauthor{\bsnm{Dalla~Mura}, \binits{M.}},
\oauthor{\bsnm{Janex}, \binits{G.}},
\oauthor{\bsnm{Socquet}, \binits{A.}}:
Seismic source characterization from gnss data using deep learning.
Authorea Preprints
(2022)
\end{botherref}
\endbibitem

\bibitem{Rogers2003}
\begin{barticle}
\bauthor{\bsnm{Rogers}, \binits{G.}},
\bauthor{\bsnm{Dragert}, \binits{H.}}:
\batitle{Episodic tremor and slip on the cascadia subduction zone: The chatter
  of silent slip}.
\bjtitle{Science}
\bvolume{300},
\bfpage{1942}--\blpage{1943}
(\byear{2003}).
\doiurl{10.1126/SCIENCE.1084783/ASSET/DCC117CE-4CAD-4799-90C0-B9DFE257EA05/ASSETS/GRAPHIC/SE2431617002.JPEG}
\end{barticle}
\endbibitem

\bibitem{Wech2010}
\begin{barticle}
\bauthor{\bsnm{Wech}, \binits{A.G.}}:
\batitle{Interactive tremor monitoring}.
\bjtitle{Seismological Research Letters}
\bvolume{81},
\bfpage{664}--\blpage{669}
(\byear{2010}).
\doiurl{10.1785/GSSRL.81.4.664}
\end{barticle}
\endbibitem

\bibitem{hayes2018slab2}
\begin{barticle}
\bauthor{\bsnm{Hayes}, \binits{G.P.}},
\bauthor{\bsnm{Moore}, \binits{G.L.}},
\bauthor{\bsnm{Portner}, \binits{D.E.}},
\bauthor{\bsnm{Hearne}, \binits{M.}},
\bauthor{\bsnm{Flamme}, \binits{H.}},
\bauthor{\bsnm{Furtney}, \binits{M.}},
\bauthor{\bsnm{Smoczyk}, \binits{G.M.}}:
\batitle{Slab2, a comprehensive subduction zone geometry model}.
\bjtitle{Science}
\bvolume{362}(\bissue{6410}),
\bfpage{58}--\blpage{61}
(\byear{2018})
\end{barticle}
\endbibitem

\bibitem{gao2012scaling}
\begin{barticle}
\bauthor{\bsnm{Gao}, \binits{H.}},
\bauthor{\bsnm{Schmidt}, \binits{D.A.}},
\bauthor{\bsnm{Weldon}, \binits{R.J.}}:
\batitle{Scaling relationships of source parameters for slow slip events}.
\bjtitle{Bulletin of the Seismological Society of America}
\bvolume{102}(\bissue{1}),
\bfpage{352}--\blpage{360}
(\byear{2012})
\end{barticle}
\endbibitem

\bibitem{okada1985surface}
\begin{barticle}
\bauthor{\bsnm{Okada}, \binits{Y.}}:
\batitle{Surface deformation due to shear and tensile faults in a half-space}.
\bjtitle{Bulletin of the seismological society of America}
\bvolume{75}(\bissue{4}),
\bfpage{1135}--\blpage{1154}
(\byear{1985})
\end{barticle}
\endbibitem

\bibitem{lecun2015deep}
\begin{barticle}
\bauthor{\bsnm{LeCun}, \binits{Y.}},
\bauthor{\bsnm{Bengio}, \binits{Y.}},
\bauthor{\bsnm{Hinton}, \binits{G.}}:
\batitle{Deep learning}.
\bjtitle{nature}
\bvolume{521}(\bissue{7553}),
\bfpage{436}--\blpage{444}
(\byear{2015})
\end{barticle}
\endbibitem

\bibitem{vaswani2017attention}
\begin{botherref}
\oauthor{\bsnm{Vaswani}, \binits{A.}},
\oauthor{\bsnm{Shazeer}, \binits{N.}},
\oauthor{\bsnm{Parmar}, \binits{N.}},
\oauthor{\bsnm{Uszkoreit}, \binits{J.}},
\oauthor{\bsnm{Jones}, \binits{L.}},
\oauthor{\bsnm{Gomez}, \binits{A.N.}},
\oauthor{\bsnm{Kaiser}, \binits{{\L}.}},
\oauthor{\bsnm{Polosukhin}, \binits{I.}}:
Attention is all you need.
Advances in neural information processing systems
\textbf{30}
(2017)
\end{botherref}
\endbibitem

\bibitem{ide2012variety}
\begin{botherref}
\oauthor{\bsnm{Ide}, \binits{S.}}:
Variety and spatial heterogeneity of tectonic tremor worldwide.
Journal of Geophysical Research: Solid Earth
\textbf{117}(B3)
(2012)
\end{botherref}
\endbibitem

\bibitem{Itoh2022}
\begin{barticle}
\bauthor{\bsnm{Itoh}, \binits{Y.}},
\bauthor{\bsnm{Aoki}, \binits{Y.}},
\bauthor{\bsnm{Fukuda}, \binits{J.}}:
\batitle{Imaging evolution of cascadia slow-slip event using high-rate gps}.
\bjtitle{Scientific Reports 2022 12:1}
\bvolume{12},
\bfpage{1}--\blpage{12}
(\byear{2022}).
\doiurl{10.1038/s41598-022-10957-8}
\end{barticle}
\endbibitem

\bibitem{bletery2020slip}
\begin{barticle}
\bauthor{\bsnm{Bletery}, \binits{Q.}},
\bauthor{\bsnm{Nocquet}, \binits{J.-M.}}:
\batitle{Slip bursts during coalescence of slow slip events in cascadia}.
\bjtitle{Nature communications}
\bvolume{11}(\bissue{1}),
\bfpage{2159}
(\byear{2020})
\end{barticle}
\endbibitem

\bibitem{licciardi2022instantaneous}
\begin{barticle}
\bauthor{\bsnm{Licciardi}, \binits{A.}},
\bauthor{\bsnm{Bletery}, \binits{Q.}},
\bauthor{\bsnm{Rouet-Leduc}, \binits{B.}},
\bauthor{\bsnm{Ampuero}, \binits{J.-P.}},
\bauthor{\bsnm{Juhel}, \binits{K.}}:
\batitle{Instantaneous tracking of earthquake growth with elastogravity
  signals}.
\bjtitle{Nature}
\bvolume{606}(\bissue{7913}),
\bfpage{319}--\blpage{324}
(\byear{2022})
\end{barticle}
\endbibitem

\bibitem{schreiber2000surrogate}
\begin{barticle}
\bauthor{\bsnm{Schreiber}, \binits{T.}},
\bauthor{\bsnm{Schmitz}, \binits{A.}}:
\batitle{Surrogate time series}.
\bjtitle{Physica D: Nonlinear Phenomena}
\bvolume{142}(\bissue{3-4}),
\bfpage{346}--\blpage{382}
(\byear{2000})
\end{barticle}
\endbibitem

\bibitem{lay1995modern}
\begin{bbook}
\bauthor{\bsnm{Lay}, \binits{T.}},
\bauthor{\bsnm{Wallace}, \binits{T.C.}}:
\bbtitle{Modern Global Seismology}.
\bpublisher{Elsevier}, \blocation{???}
(\byear{1995})
\end{bbook}
\endbibitem

\bibitem{ioffe2015batch}
\begin{bchapter}
\bauthor{\bsnm{Ioffe}, \binits{S.}},
\bauthor{\bsnm{Szegedy}, \binits{C.}}:
\bctitle{Batch normalization: Accelerating deep network training by reducing
  internal covariate shift}.
In: \bbtitle{International Conference on Machine Learning},
pp. \bfpage{448}--\blpage{456}
(\byear{2015}).
\bcomment{pmlr}
\end{bchapter}
\endbibitem

\bibitem{agarap2018deep}
\begin{botherref}
\oauthor{\bsnm{Agarap}, \binits{A.F.}}:
Deep learning using rectified linear units (relu).
arXiv preprint arXiv:1803.08375
(2018)
\end{botherref}
\endbibitem

\bibitem{srivastava2014dropout}
\begin{barticle}
\bauthor{\bsnm{Srivastava}, \binits{N.}},
\bauthor{\bsnm{Hinton}, \binits{G.}},
\bauthor{\bsnm{Krizhevsky}, \binits{A.}},
\bauthor{\bsnm{Sutskever}, \binits{I.}},
\bauthor{\bsnm{Salakhutdinov}, \binits{R.}}:
\batitle{Dropout: a simple way to prevent neural networks from overfitting}.
\bjtitle{The journal of machine learning research}
\bvolume{15}(\bissue{1}),
\bfpage{1929}--\blpage{1958}
(\byear{2014})
\end{barticle}
\endbibitem

\bibitem{bottou2018optimization}
\begin{barticle}
\bauthor{\bsnm{Bottou}, \binits{L.}},
\bauthor{\bsnm{Curtis}, \binits{F.E.}},
\bauthor{\bsnm{Nocedal}, \binits{J.}}:
\batitle{Optimization methods for large-scale machine learning}.
\bjtitle{Siam Review}
\bvolume{60}(\bissue{2}),
\bfpage{223}--\blpage{311}
(\byear{2018})
\end{barticle}
\endbibitem

\bibitem{kingma2014adam}
\begin{botherref}
\oauthor{\bsnm{Kingma}, \binits{D.P.}},
\oauthor{\bsnm{Ba}, \binits{J.}}:
Adam: A method for stochastic optimization.
arXiv preprint arXiv:1412.6980
(2014)
\end{botherref}
\endbibitem

\bibitem{he2015delving}
\begin{bchapter}
\bauthor{\bsnm{He}, \binits{K.}},
\bauthor{\bsnm{Zhang}, \binits{X.}},
\bauthor{\bsnm{Ren}, \binits{S.}},
\bauthor{\bsnm{Sun}, \binits{J.}}:
\bctitle{Delving deep into rectifiers: Surpassing human-level performance on
  imagenet classification}.
In: \bbtitle{Proceedings of the IEEE International Conference on Computer
  Vision},
pp. \bfpage{1026}--\blpage{1034}
(\byear{2015})
\end{bchapter}
\endbibitem

\bibitem{chollet2015keras}
\begin{botherref}
\oauthor{\bsnm{Chollet}, \binits{F.}}, et al.:
Keras.
\url{https://github.com/fchollet/keras}
\end{botherref}
\endbibitem

\bibitem{abadi2016tensorflow}
\begin{botherref}
\oauthor{\bsnm{Abadi}, \binits{M.}},
\oauthor{\bsnm{Agarwal}, \binits{A.}},
\oauthor{\bsnm{Barham}, \binits{P.}},
\oauthor{\bsnm{Brevdo}, \binits{E.}},
\oauthor{\bsnm{Chen}, \binits{Z.}},
\oauthor{\bsnm{Citro}, \binits{C.}},
\oauthor{\bsnm{Corrado}, \binits{G.S.}},
\oauthor{\bsnm{Davis}, \binits{A.}},
\oauthor{\bsnm{Dean}, \binits{J.}},
\oauthor{\bsnm{Devin}, \binits{M.}}, et al.:
Tensorflow: Large-scale machine learning on heterogeneous distributed systems.
arXiv preprint arXiv:1603.04467
(2016)
\end{botherref}
\endbibitem

\bibitem{2020SciPy-NMeth}
\begin{barticle}
\bauthor{\bsnm{Virtanen}, \binits{P.}},
\bauthor{\bsnm{Gommers}, \binits{R.}},
\bauthor{\bsnm{Oliphant}, \binits{T.E.}},
\bauthor{\bsnm{Haberland}, \binits{M.}},
\bauthor{\bsnm{Reddy}, \binits{T.}},
\bauthor{\bsnm{Cournapeau}, \binits{D.}},
\bauthor{\bsnm{Burovski}, \binits{E.}},
\bauthor{\bsnm{Peterson}, \binits{P.}},
\bauthor{\bsnm{Weckesser}, \binits{W.}},
\bauthor{\bsnm{Bright}, \binits{J.}},
\bauthor{\bsnm{{van der Walt}}, \binits{S.J.}},
\bauthor{\bsnm{Brett}, \binits{M.}},
\bauthor{\bsnm{Wilson}, \binits{J.}},
\bauthor{\bsnm{Millman}, \binits{K.J.}},
\bauthor{\bsnm{Mayorov}, \binits{N.}},
\bauthor{\bsnm{Nelson}, \binits{A.R.J.}},
\bauthor{\bsnm{Jones}, \binits{E.}},
\bauthor{\bsnm{Kern}, \binits{R.}},
\bauthor{\bsnm{Larson}, \binits{E.}},
\bauthor{\bsnm{Carey}, \binits{C.J.}},
\bauthor{\bsnm{Polat}, \binits{{\. I}.}},
\bauthor{\bsnm{Feng}, \binits{Y.}},
\bauthor{\bsnm{Moore}, \binits{E.W.}},
\bauthor{\bsnm{{VanderPlas}}, \binits{J.}},
\bauthor{\bsnm{Laxalde}, \binits{D.}},
\bauthor{\bsnm{Perktold}, \binits{J.}},
\bauthor{\bsnm{Cimrman}, \binits{R.}},
\bauthor{\bsnm{Henriksen}, \binits{I.}},
\bauthor{\bsnm{Quintero}, \binits{E.A.}},
\bauthor{\bsnm{Harris}, \binits{C.R.}},
\bauthor{\bsnm{Archibald}, \binits{A.M.}},
\bauthor{\bsnm{Ribeiro}, \binits{A.H.}},
\bauthor{\bsnm{Pedregosa}, \binits{F.}},
\bauthor{\bsnm{{van Mulbregt}}, \binits{P.}},
\bauthor{\bsnm{{SciPy 1.0 Contributors}}}:
\batitle{{{SciPy} 1.0: Fundamental Algorithms for Scientific Computing in
  Python}}.
\bjtitle{Nature Methods}
\bvolume{17},
\bfpage{261}--\blpage{272}
(\byear{2020}).
\doiurl{10.1038/s41592-019-0686-2}
\end{barticle}
\endbibitem

\end{thebibliography}


\end{document}